\documentstyle[emulateapj,psfig]{article}
\begin{document}

   \title{The Peak Energy Distribution of  the $\nu F_{\nu}$ Spectra and the Implications for the Jet Structure Models of Gamma-Ray Bursts
}

\author{E. W. Liang$^{1,2,3}$ and  Z. G. Dai$^{1}$}
\affil{$^1$Department of Astronomy, Nanjing University, Nanjing 210093, China;
ewliang@nju.edu.cn.\\ $^2$Department of Physics, Guangxi University, Nanning 530004, China.\\
$^3$National Astronomical Observatories/Yunnan Observatory, Chinese Academy of Sciences, Kunming
650011, China.}

\begin{abstract}
We study the peak energy ($E_{\rm{p}}$) distribution of the $\nu F_{\nu}$ spectra of gamma-ray
bursts (GRBs) and X-ray flashes (XRFs) with a sample of 57 bursts observed by {\em High Energy
Transient Explorer 2} ({\em HETE-2}) French Gamma Telescope and discuss its implications for the
jet structure models. Combining the observed $E_{\rm{p}}$ distribution of {\em HETE--2} GRBs/XRFs
with that of BATSE GRBs, we find that the observed $E_{\rm{p}}$ distribution of GRBs/XRFs is a
bimodal one with peaks of $\lesssim 30$ keV and $\sim 160-250$ keV. According to the
recently-discovered equivalent-isotropic energy-$E_{\rm{p}}$ relationship, such a bimodal
distribution implies a two-component structure of GRB/XRF jets. A simple simulation analysis shows
that this structured jet model does roughly reproduce a bimodal distribution with peaks of $\sim
15$ and $\sim 200$ keV. We argue that future observations of  the peak of $\sim 15$ keV in the
$E_{\rm{p}}$ distribution would be evidence supporting this model. {\em Swift}, which covers an
energy band of 0.2--150 keV, is expected to provide a key test for our results.
\end{abstract}

\section{Introduction}               
X-ray flashes (XRFs) have being gotten a lot of attention in the last 2 years (Heise et al. 2001;
Kippen et al. 2003). They are thought to be a lower energy extension of the known gamma-ray burst
(GRB) population, based on the fact that their spectral behaviors are similar to those of GRBs
(Kippen et al. 2003; Barraud et al. 2003; Sakamoto et al. 2004; Lamb et al. 2003a, b, 2004). The
nature of a narrow cluster of the observed $E_{\rm{p}}$ distribution of BATSE GRBs remains poorly
understood, which might be related to the jet structure of GRBs. XRFs broaden the energy coverage
of prompt GRB emission and may bring more signatures of the jet structure of GRBs (Lamb et al.
2003a, b, 2004).

The jet structure models are currently under heavy debate. Any model should present a unified
description for GRBs and XRFs. Two currently competing models are the structured jet model
(M\'{e}sz\'{a}ros, Rees \& Wijers 1998; Dai \& Gou 2001; Rossi, Lazzati, \& Rees 2002; Zhang \&
M\'{e}sz\'{a}ros 2002; Granot \& Kumar 2003; Kumar and Granot 2003; Panaitescu \& Kumar 2003; Wei
\& Jin 2003) and the uniform model (e.g., Rhoads 1999; Frail et al. 2001). Zhang et al. (2004) show
that the current GRB/XRF prompt emission/afterglow data can be described by a quasi--Gaussian-type
(or similar structure) structured jet with a typical opening angle of $\sim 6^{o}$ and with a
standard jet energy of $\sim 10^{51}$ ergs. Alternatively, based on the {\em High Energy Transient
Explorer 2} ({\em HETE-2}) observations, Lamb et al. (2003a, b, 2004) propose that the uniform jet
model can reasonably describe the unified scheme of GRBs/XRFs. Very recently, the two-component jet
model was advocated by Berger et al. (2003) based on observations of GRB 030329, which has two
different jet breaks in an early optical afterglow light curve (0.55 days, Price et al. 2003) and
in a late radio light curve (9.8 days). Millimeter observations of this burst further support the
two--component jet model (Sheth et al. 2003). Numerical calculations of such a model were performed
by Huang et al. (2004 ). This model suggests that a GRB/XRF jet has two components: a narrow,
highly relativistic one, and a wide, mildly relativistic one. When the line of sight of an observer
is within the narrow component, the observed burst is a typical GRB, but when the line of sight is
pointing to the wide--component, it is an XRF.

A broad spectral energy distribution could constrain the jet structure models. A low peak energy of
the $\nu F_{\nu}$ spectrum ($E_{\rm{p}}<50$ keV) and weak gamma--ray fluxes ($F<0.2$ photons
cm$^{-2}$ s$^{-1}$, 50--300 keV energy range) distinguish XRFs from typical GRBs (Kippen et al.
2003; Mochkovitch et al. 2003). It is well known that the observed $E_{\rm{p}}$ distribution of
BATSE GRBs is narrowly clustered. Does the observed $E_{\rm{p}}$ distribution of XRFs exhibit a
similar feature? In this Letter, we focus on this question. We analyze the observed $E_{\rm{p}}$
distribution with a sample of 57 bursts observed by {\em HETE-2}/ French Gamma Telescope (FREGATE).
Combining the observed $E_{\rm{p}}$ distribution of {\em HETE-2} GRBs/XRFs with that of BATSE GRBs,
we find that the observed $E_{\rm{p}}$ distribution of GRBs/XRFs is a bimodal one peaking at
$\lesssim 30$ keV and $\sim 160-250$ keV. With respect to this result, we suggest that the
two--component jet model is a reasonable candidate model for GRB/XRF jets. A simulation analysis
confirms this suggestion.

\section{Distribution of $E_{\rm{p}}$}
We make a search for {\em HETE-2} GRBs/XRFs reported in literature and on the {\em HETE-2} Web
site\footnote{http://space.mit.edu/HETE/Bursts/}. All the bursts with $E_{\rm{p}}$ or fluences
($S$) in the available energy bands 7--30 keV and 30--400 keV are included in our sample. We obtain
a sample that includes 57 bursts. Among them, 49 of the bursts are taken from Barraud et al.
(2003), Atteia (2003), Sakamoto et al. (2004), Lamb et al. (2003a, b, 2004), and the {\em HETE--2}
Web site. Their $E_{\rm{p}}$ values are derived from spectral fittings. Please note that the
$E_{\rm{p}}$ values of GRB 010923, 011216, and 021004 presented in Barraud et al. (2003) are
incorrect, and they are taken from Lamb et al. (2003a, b, 2004). For the other eight bursts, GRB
030824, 030823, 030725, 030913, 030528, 030519, 030418, and 030416, only fluences in the energy
bands of 7--30 keV and 30--400 keV are available. For these bursts, we estimate their $E_{\rm{p}}$
by their spectral hardness ratios, which are defined as $R=S_{30-400\ {\rm{keV}}}/S_{7-30\
{\rm{keV}}}$. Since the spectra of GRBs/XRFs can be well fitted by the Band function (Band et al.
1993) with similar spectral indices (Kippen et al. 2003; Barraud et al. 2003), their $E_{\rm{p}}$
should be proportional to $R$. A best fit to the data presented in Barraud et al. (2003) derives
$\log E_{\rm{p}}=(1.52\pm 0.05)+(0.92\pm 0.07)\log R$ with a linear coefficient of 0.93 and a
chance probability $p<0.0001$ ($N=32$, without considering GRB 010923, 011216, or 021004). We thus
estimate the $E_{\rm{p}}$ values of the above eight bursts by using this relation.

We show the $E_{\rm{p}}$ distribution in a range of $\log E_{\rm{p}}/{\rm{keV}}=0.6-3.0$ with a
step of 0.23 for these bursts in Figure 1a. It is found that the distribution has three peaks at
30, 160, and 450 keV. We note that the peaks of 160 and 450 keV seem to be embedded in one peak,
and the gap at $E_{\rm{p}}=275$ keV is likely to be fake. The spectral analysis for a bright BATSE
GRB sample by Preece et al. (2000) has shown that the $E_{\rm{p}}$ values are clustered at 100-1000
keV with a peak of $\sim 250$ keV (the dotted line in panel (a) of Figure 1). We thus suspect that
the peaks of 160 keV and 450 keV are likely to be embedded in one peak which is similar to that of
BATSE GRB sample. If the case really shows one peak, the $E_{\rm{p}}$ distributions observed by
{\em HETE--2} and by BATSE in the range of $100-1000$ keV should be consistent. We examine this
hypothesis by a Kolmogorov-Smirnoff (K-S) test (Press et al. 1997, p.617). The result of the K-S
test is described by a statistic of $P_{\rm{K-S}}$: a small value of $P_{\rm{K-S}}$ indicates a
significant difference between two distributions ($P_{\rm{K-S}}=1$ indicates that two distributions
are identical, and $P_{\rm{K-S}}<0.0001$ suggests that the consistency of two distributions should
be rejected; e.g., Bloom 2003). We obtain $P_{\rm{K-S}}=0.22$, indicating that the consistency of
the two distributions is acceptable. However, their difference is still quite significant. This
difference might be due to a strong sample selection effect in the BATSE GRB sample presented by
Preece et al. (2000), who considered only those bursts with total fluence $\geq 5\times 10^{-5}$
ergs cm$^{-2}$ or peak fluxes higher than 10 photons cm $^{-2}$ s$^{-1}$ in a 1.024 s timescale. To
avoid such a sample selection effect, we further compare the distributions of the hardness ratios
of {\em HETE-2} bursts and BATSE bursts in Figure 1b. In Figure 1b, the BATSE GRB sample includes
all of the long-duration bursts without any sample selection effect (1213 events, from BATSE
Current Catalog). A K--S test to the two distributions in the range of $\log R=0.3-1.5$ derives
$P_{\rm{KS}}=0.95$, strongly suggesting a consistency between the two distributions in that range.
Thus, we suggest that the $E_{\rm{p}}$ distribution in 100--1000 keV should form one sole peak,
centering at $\sim 160-250$ keV.

The peak of $E_{\rm{p}}\sim 30$ keV or $R\sim 1$ seems to be a unique one. A sharp cutoff occurs on
its left side. This might be caused by the limit of {\em HETE-2}. Hence, we suggest that the
$E_{\rm{p}}$ distribution should exhibit another peak of an energy $\lesssim 30$ keV.

Based on the above analysis, we propose that the $E_{\rm{p}}$ distribution of GRBs/XRFs is a
bimodal one, peaking at an energy $\lesssim 30$ keV and $\sim 160-250$ keV.

\section{Implications for the Jet structure and Unified Models of GRBs/XRFs}
The observed bimodal distribution of $E_{\rm{p}}$ for GRBs/XRFs might strongly constrain the jet
structure models of GRBs/XRFs.

From  $E_{\rm{iso,52}}\simeq [E_{\rm{p,2}}(1+z)]^{2}$, where
$E_{\rm{iso,52}}=E_{\rm{iso}}/10^{52}\rm{\ ergs}$ and $E_{\rm{p,2}}=E_{\rm{p}}/10^2{\rm{\ keV}}$
(Amati et al. 2002; Lloyd-Ronning \& Ramirez-Ruiz 2002; Atteia 2003; Sakamoto et al. 2004; Lamb et
al. 2003a, b, 2004; Liang, Dai \&Wu 2004; Yonetoku et al. 2004), and
$E_{\rm{iso,52}}(1-\cos\theta_{\rm{j}})=0.133$, where $\theta_{\rm{j}}$ is the jet opening angle
(Frail et al. 2001; Panaitescu \& Kumar 2001; Piran et al. 2001; Bloom et al. 2003; Berger,
Kulkarni, \& Frail 2003), we can derive
\begin{equation}
\theta_{\rm{j}}={\rm{arccos}} \left \{ 1-\frac{0.133}{[E_{\rm{p,2}}(1+z)]^2}\right \}.
\end{equation}

In the uniform jet model, one expects that both XRFs and GRBs should obey Eq. (1). However, this
relation cannot simply extend to any bursts with $E_{\rm{p}}(1+z)<35$ keV, because of the limit of
$\theta_{\rm{j}}<\pi/2$. The redshifts of the two extremely soft XRFs, 020903 and 030723, are 0.251
(Soderberg et al. 2003) and less than 2.1 (Fynbo et al. 2004), respectively; but their $E_{\rm{p}}$
values are less than $20$ keV. The two XRFs violate this relationship. In addition, the uniform jet
model may not accommodate the observed bimodal distribution of $E_{\rm{p}}$.

A quasi-universal Gaussian-type jet model may also present a unified picture for GRBs/XRFs.
Lloyd-Ronning, Dai, \& Zhang (2003) found that this model can reproduce the relation of the
equivalent--isotropic energy to the viewing angle, and Zhang et al. (2004a) further showed that the
current GRB/XRF prompt emission/afterglow data can be described by this model (or similarly
structured jet) with a typical opening angle of $\sim 6^{\circ}$ and with a standard jet energy of
$\sim 10^{51}$ ergs. However, the observed bimodal distribution of $E_{\rm{p}}$ is difficult to be
explained by this model.

According to the equivalent-isotropic energy--$E_{\rm{p}}$ relationship discovered recently by
Amati et al. (2002), the bimodal $E_{\rm{p}}$ distribution seems to imply a two-component structure
of GRB/XRF jets. To investigate whether or not this model can reproduce the observed bimodal
distribution of $E_{\rm{p}}$, we make a simple simulation analysis. We describe the energy per
solid angle of the two--component model by two Gaussian jets,

\begin{equation}
\epsilon=\epsilon_0(e^{-\theta_{\rm{v}}^2/2\theta_{1}^2}+\mu e^{-\theta_{\rm{v}}^2/2\theta_2^2}),
\end{equation}
where $\theta_{\rm{v}}$ is the viewing angle measured from the jet axis, $\epsilon_0$ is the
maximum value of energy per solid angle, $\mu$ is the ratio of $E_{\rm{iso}}$ in the wide component
to narrow component, and $\theta_{1}$ and $\theta_{2}$ are characteristic angular widthes of the
narrow and wide components, respectively. Since $E_{\rm{p}} \propto \epsilon^{0.5}$, the observed
$E_{\rm{p}}$ should be given by

\begin{equation}
E_{\rm{p}}=E_{\rm{p,\ 0}}(1+z)(e^{-\theta_{\rm{v}}^2/2\theta_{1}^2}+\mu
e^{-\theta_{\rm{v}}^2/2\theta_2^2})^{1/2}.
\end{equation}

Similar to Lloyd-Ronning, Dai, \& Zhang (2003) and Zhang et al. (2004), we assume that the two
components are quasi-universal, where ``quasi" means that the parameters of this model have a
dispersion but are not invariable. We perform a simple Monte Carlo simulation analysis with the
distributions of these parameters. The probability of observing a GRB/XRF with $\theta_{\rm{v}}$ is
proportional to $\sin \theta_{\rm{v}}$. One can expect this probability to be random. Thus, we
assume that $\sin\theta_{\rm{v}}$ is uniformly distributed in the range of 0--1. The $E_{\rm{p,\
0}}$ distribution should be mainly determined by a bright GRB sample. Since the observed
$E_{\rm{p}}$ for bright BATSE GRBs are narrowly clustered at $200-400$ keV  and since the measured
redshift distribution is around $1$, we take the differential distribution of $E_{\rm{p,\ 0}}$ as
that of $E_{\rm{p}}$ for the bright GRBs, but centered at $\log E_{\rm{p,0}}=2.80$ (i.e.,
$E_{\rm{p,0}}=630$ keV), which is given by $w(\log E_{\rm{p,0}})=0.018 \exp \{-2[(\log
E_{\rm{p,0}}-2.80)^2]/0.45^2\}$, where the coefficient 0.018 is a normalized constant. We assume
that the redshift distribution is the same as the one of Bloom (2003), who assumed that the burst
rate as a function of redshift is proportional to the star formation rate, and who presented the
observed redshift distribution incorporating observational biases (model SF1 from Porciani \& Madau
2001 is used in this work). We also restrict $z\leq4.5$ because the largest $z$ is 4.5 in our
present GRB sample. For $\theta_1$ and $\mu$, we cannot reasonably model their distributions with
the present data, and thus we simply estimate their values as follows. Since the mean value of the
jet opening angles of 16 GRBs presented in Bloom et al. (2003) is $\sim 0.15$ rad (without
considering the eight GRBs whose limits of jet opening angles are presented), we take $\theta_1\sim
0.15$ rad. Based on the results shown in Figure 1, we have $\mu=E_{\rm{iso,XRF}}/E_{\rm{iso,GRB}}
\simeq 10^{-1.7}$. The $\theta_2$ is the most poorly understood among these parameters. We let it
be an adjustable variable with a limit of $\theta_2>\theta_1$. In our simulation analysis, we take
$\theta_2=0.32$ rad (see below).

We simulate a sample of $10^5$ GRBs/XRFs. Our simulation analysis procedure is described as
follows. To derive a value of parameter $x$ for a given burst ($x$ is one of $E_{\rm{p,0}}$, $z$,
and $\theta_v$), we first derive the accumulative probability distributions of these parameters
$P(x)$ ($0<P(x)\leq 1$), then generate a random number $m$ ($0<m\leq 1$), and finally obtain the
value of $x$ from the inverse function of $P(x)=m$; i.e., $x=P^{-1}(m)$. The values of $\theta_1$
and $\mu$ are fixed at 0.15 rad and $10^{-1.7}$, respectively.  The value of $\theta_2$ is an
adjustable variable with a limit of $\theta_2>\theta_1$. We find that $\theta_2=0.32$ rad can
roughly reproduce the $E_{\rm{p}}$ distribution shown in Figure 1. We calculate the $E_{\rm{p}}$
for each simulated GRB/XRF with the above parameters using Eq. (3). The $E_{\rm{p}}$ distribution
is shown in Figure 2. We find that the distribution is bimodal with peaks of $\sim 15$ and $\sim
200$ keV and with a valley at $\sim 50$ keV. These results show that the two-Gaussian jet model can
roughly reproduce the bimodal distribution of the observed $E_{\rm{p}}$.

In our simulation, we do not consider any instrument threshold setting. The energy bandpass of {\em
HETE--2}/FREGATE is 7--400 keV. From Figure 1, we find a sharp cutoff at $\log
E_{\rm{p}}/{\rm{keV}}=1.3$ (i.e., $E_{\rm{p}}=20$ keV), which is close to the lowest end of the
{\em HETE--2} energy bandpass. This $E_{\rm{p}}$ value might reflect the effective threshold of
{\em HETE--2}. We roughly estimate the ratio of observable GRBs to XRFs for {\em HETE--2} with this
threshold in our simulation analysis, and find that this ratio is about 2.2:1. This is in a good
agreement with {\em HETE--2}/FREGATE observations (39 {\em HETE--2} GRBs and 18 XRFs in the {\em
HETE--2}/FREGATE sample).

\section{Conclusions and Discussion}
We have studied the observed $E_{\rm{p}}$ distribution of 57 {\em HETE-2}/FREGATE bursts, and
discuss its implications for the jet structure models. Combining the observed $E_{\rm{p}}$
distribution of {\em HETE-2} GRBs/XRFs with that of BATSE GRBs, we suggest that the observed
$E_{\rm{p}}$ distribution of GRBs/XRFs is bimodal with peaks of $\lesssim 30$ keV and $\sim
160-250$ keV. According to the recently--discovered equivalent-isotropic energy--$E_{\rm{p}}$
relationship, we find that the bimodal distribution can be explained by the two--component model of
GRB/XRF jets. A simple simulation analysis shows that this structured jet model does roughly
reproduce the bimodal distribution with peaks of $\sim 15$ keV and $\sim 200$ keV.

The peak of $\sim 15$ keV in the simulated $E_{\rm{p}}$ distribution is key evidence for the two--
component jet model. It is near the lowest end of the energy bandpass of {\em HETE-2}/FREGATE.
Fortunately, {\em HETE-2} provides a weak clue to this peak. A more sensitive instrument than {\em
HETE--2} with an energy bandpass $1-50$ keV is required to further confirm this peak. {\em Swift},
which covers an energy band of 0.2--150 keV (we mark this region in Figure 2 with diagonal
lines)\footnote{ http://swift.gsfc.nasa.gov/science/instruments/} is expected to provide a key test
for it.

Simulations of the propagation and eruption of relativistic jets in massive Wolf-Rayet stars by
Zhang, Woosley, \& Heger (2004b) show that an erupting jet has a highly relativistic, strongly
collimated core, and a moderately relativistic, less energetic cocoon. The cocoon expands and
becomes visible at larger angles. The energy ratio of the cocoon to the core in their simulation is
about one order. From our simulation results, we find that it is $\sim
(E_{\rm{p,GRB}}/E_{\rm{p,XRF}})^2(\theta_{1}/\theta_2)^2\sim 40$, being roughly consistent with
their results. Their simulations seem to support the two-component jet model. We have noted that
the ability of the cocoon to cause an XRF depends sensitively on its Lorentz factor, which is
determined by the degree of mixing between the jet and envelope material. Matzner (2003) argued
that this mixing might be difficult to resolve in numerical simulations.

A two-component jet was suggested to be universal for GRB/XRF phenomena in this Letter, based on
the multi-wavelength observations of GRB 0303029 (Berger et al. 2003; Sheth et al. 2003) and the
bimodal distribution of $E_{\rm {p}}$. It should be pointed out that other jet models such as
uniform jets and single-component-universal jets were proposed to explain numerous observations on
the afterglows and some correlations (e.g., Lamb et al. 2003b; Lloyd-Ronning et al. 2004;
Lloyd-Ronning \& Zhang 2004). Thus, one would expect strong evidence showing which jet model is
more reasonable.

We are very grateful to the anonymous referee for his/her valuable suggestions. We also sincerely
thank Bing Zhang and Johan P.U. Fynbo for their helpful comments. These suggestions and comments
have enabled us to improve  the manuscript greatly. This work is supported by the National Natural
Science Foundation of China (grants 10233010 and 10221001), the National 973 Project (NKBRSF
G19990754), the Natural Science Foundation of Yunnan (2001A0025Q), and the Research Foundation of
Guangxi University.

\begin{figure}
\plotone{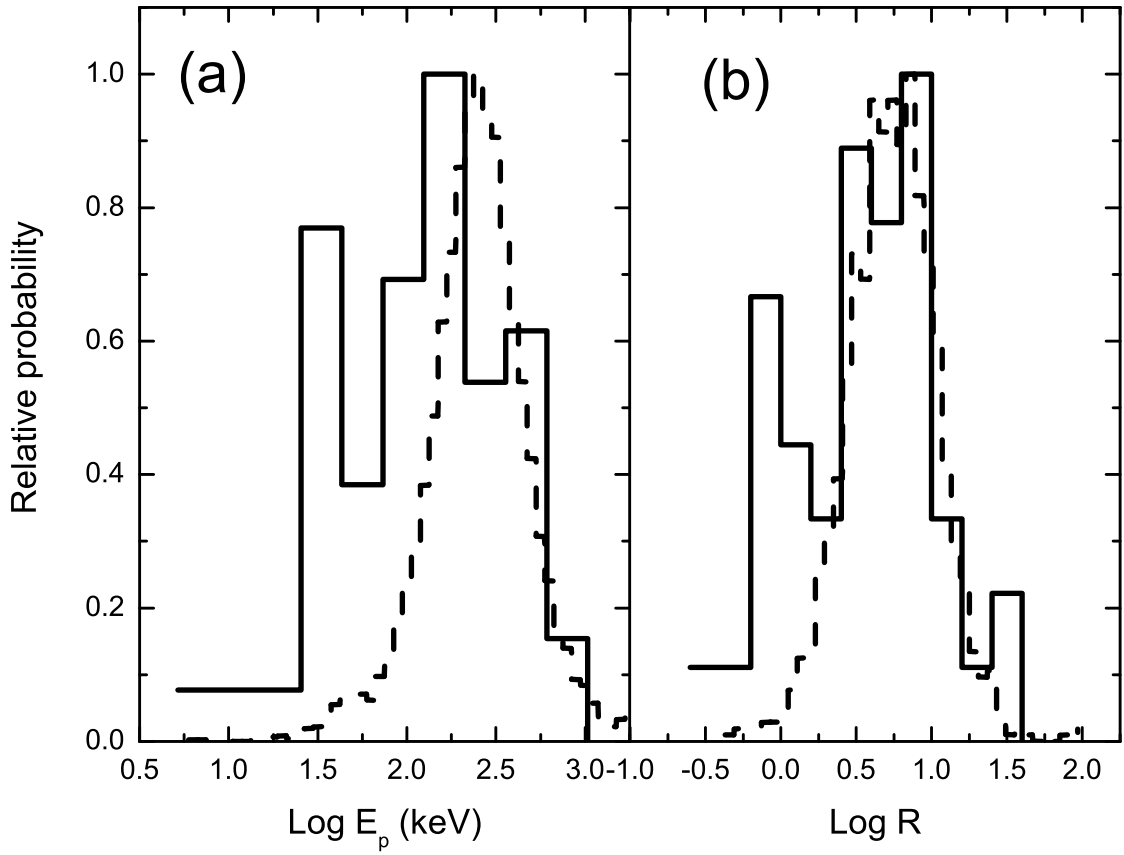} \caption{Observed $E_{\rm{p}}$ [panel (a)] and hardness ratio [panel (b) ]
distributions of {em HETE-2}/FREGATE GRBs/XRFs. In panel (a), the dashed line is the observed
$E_{\rm{p}}$ distribution of a bright BATSE GRB sample taken from Preece et al. (2000). In panel
(b), the dashed line is the observed hardness ratio distribution of all long-duration BATSE GRBs
without any sample selection effect (from BATSE Current Catalog).\label{fig1}}
\end{figure}

\begin{figure}
\plotone{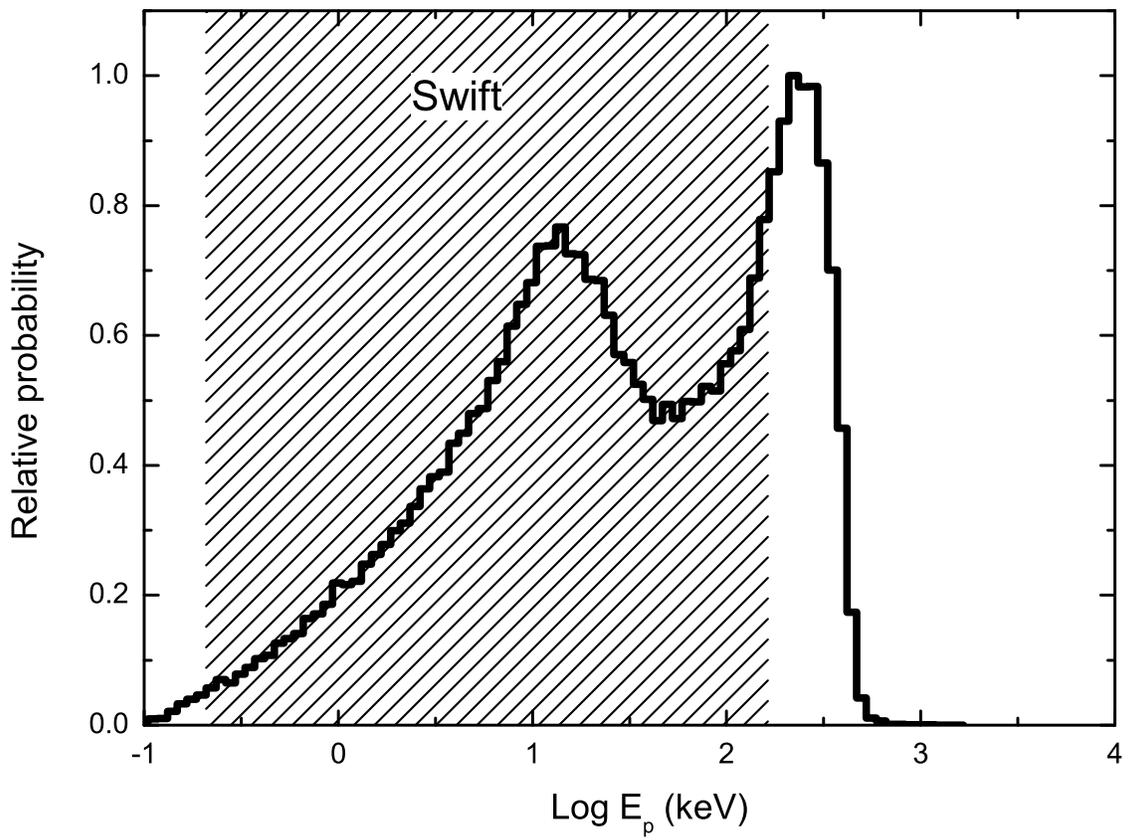} \caption{Simulated $E_{\rm{p}}$ distribution of the two-quasi-universal Gaussian
jet model. The  diagonal line region is the energy band of {\em Swift}. \label{fig2}}
\end{figure}

\end{document}